# Nonlinear Disorder Mapping Through Three-Wave Mixing

A. Pasquazi
A. Busacca
S. Stivala
R. Morandotti
G. Assanto

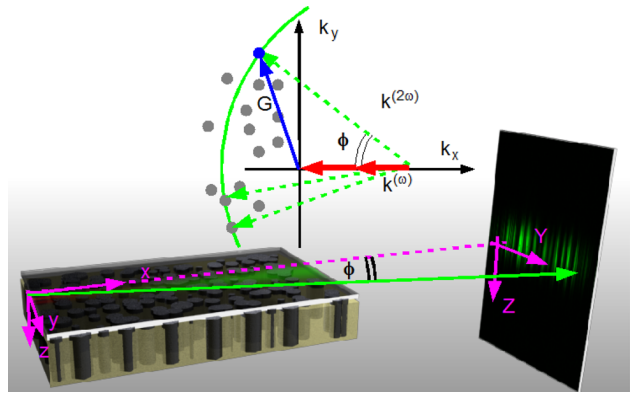

**Abstract:** We implement a simple and powerful approach to characterize the domain distribution in the bulk of quadratic ferroelectric crystals via far-field second-harmonic spectroscopy. The approach is demonstrated in a lithium tantalate sample with periodic electric field poling and random mark-to-space ratio.

## 1. Introduction

Wave propagation in nonhomogeneous and disordered media has been intensively studied in a number of optical systems for both single and multiple scattering dynamics. Following the recent progress in the poling technology of ferroelectric crystals, a number of studies have dealt with nonlinear frequency conversion in nonhomogeneous quadratic dielectrics, ranging from nonlinear lattices [1]–[5] to disordered media [6]–[12]. Such investigations have been triggered not only by the fundamental interest in understanding the interplay between wave interaction and the spatial distribution associated to nonlinearity but also by their potential impact on applications aiming at broadband frequency generation or conversion by three-wave mixing (TWM) [6]–[9], [12]. Among other effects, since momentum conservation in quadratic TWM enhances the interaction, multidimensional nonlinear lattices have been proven to yield signal generation along various directions of propagation [13], both in free wave [1], [2] and self-guided [4], [14] regimes.

In this paper, we address second-harmonic generation (SHG) in quadratic crystals with a nonhomogeneous distribution of the ferroelectric domain orientations, i.e., of the sign of the nonlinearity. As pointed out in some pioneering papers [15], [16] and in more recent work [2], [13], [17], [18], the generated second harmonic (SH) inherently possesses the spectral properties of the nonhomogeneous quadratic nonlinearity: phase matching maps the momenta of the spatial degenerate-TWM nonlinearity into SH wave vectors.

Hereby, we exploit the formal equivalence of the SH far field with the Fourier transform of the spatial distribution of the quadratic nonlinearity [16] in order to study a random distribution of ferroelectric domains in quadratic crystals. Specifically, we use this fully optical approach to reconstruct the Fourier spectrum of an arbitrary distribution of ferroelectric domains in a 1-cm-long periodically electric-field-poled LiTaO$_3$ sample designed for SHG via Quasi Phase Matching (QPM), by carrying out a set of SHG experiments and analyzing both role and properties of the random component of the mark-to-space ratio of its QPM grating. The reconstructed distribution of the quadratic coefficient inside the crystal is then used to numerically reproduce the data, obtaining quite a good agreement with the experimental results.

## 2. Theory

The Fourier transformation linking the SH far field with the ferroelectric domain distribution in an $e-ee$ degenerate-TWM interaction can be illustrated with reference to standard type-I SHG in the undepleted pump regime. [19] The SH electric field $\mathcal{E}^{(2\omega)}(t) = Re\{E^{(2\omega)}e^{-i2\omega t}\}$ generated by a fundamental-frequency (FF) field $\mathcal{E}^{(\omega)}(t) = Re\{E^{(\omega)}e^{-i\omega t}\}$ is ruled by

$$\nabla^2 E^{(2\omega)}(\mathbf{r}) + k^{(2\omega)} E^{(2\omega)}(\mathbf{r}) = -\left(k_0^{(2\omega)}\right)^2 \chi^{(2)}(\mathbf{r}) \left[E^{(\omega)}(\mathbf{r})\right]^2 \quad (1)$$

where $k_0^{(2\omega)}$ and $k^{(2\omega)}$ are the wavenumbers in vacuum and in the medium, respectively. The nonlinear susceptibility $\chi^{(2)}$ is nonhomogeneous in space.

Considering a slowly varying envelope FF beam $E^{(\omega)}(\mathbf{r}) = E_0^{(\omega)}(\mathbf{r})e^{i\mathbf{k}^{(\omega)}\cdot\mathbf{r}}$ and using the Green function [19]

$$G(\mathbf{r},\mathbf{r}') = \frac{e^{ik^{(2\omega)}|\mathbf{r}-\mathbf{r}'|}}{|\mathbf{r}-\mathbf{r}'|} \quad (2)$$

we get

$$E^{(2\omega)}(\mathbf{r}) \propto \int G(\mathbf{r},\mathbf{r}') \chi^{(2)}(\mathbf{r}) \left[E_0^{(\omega)}(\mathbf{r})\right]^2 e^{i2\mathbf{k}_\omega \cdot \mathbf{r}'} d\mathbf{r}'. \quad (3)$$

For the sake of simplicity we ignore the refractive index discontinuities between the crystal and the surrounding environment. In the far field (for $max(|\mathbf{r}' \times \hat{\mathbf{r}}|^2/2\lambda^{(2\omega)}) \ll r$ with $\hat{\mathbf{r}}$ unit vector of $\mathbf{r}$) a Fourier transformation links the spatial distribution of the nonlinear polarizability $\propto \chi^{(2)}(\mathbf{r})[E^{(\omega)}(\mathbf{r})]^2$ with the generated SH field ($E^{(2\omega)}$)

$$E^{(2\omega)}(\mathbf{r}) \propto \frac{e^{ik^{(2\omega)}r}}{2r} \mathcal{F}\left\{\chi^{(2)}(\mathbf{r})\left[E_0^{(\omega)}(\mathbf{r})\right]^2\right\}(\Delta \mathbf{k}) \quad (4)$$

where $\mathcal{F}\{u(x)\}(y) = \int u(x)e^{ixy}dx$ is the Fourier transform and $\Delta \mathbf{k} = 2\mathbf{k}^{(\omega)} - \mathbf{k}^{(2\omega)}$ the phase mismatch. The latter can be expressed as $\Delta \mathbf{k} = 2\mathbf{k}^{(\omega)} - k^{(2\omega)}(cos(\phi)sin(\theta)\hat{\mathbf{x}} + sin(\phi)sin(\theta)\hat{\mathbf{y}} + cos(\theta)\hat{\mathbf{z}})$ with $\theta$ and $\phi$ polar angles and $z$ the polar axis in spherical coordinates. The Appendix contains a complete derivation accounting for both extraordinary- and ordinary-wave propagation in a uniaxial crystal; the Fourier relation is generally valid in the undepleted pump regime.

Eq. (4) connects the SH far field to the Fourier transform of the nonlinear coefficient $\chi^{(2)}$ in the space of phase mismatch $\Delta \mathbf{k}$, with no hypotheses on its transverse spectrum, [15], [16] not even the slowly varying envelope approximation [2], [13], [17]. Therefore, it is valid also in highly disordered media where the SH can strongly diffract due to momentum conservation [3], [10], [20].

The Fourier relation in Eq. (4) entails the noninvasive study of ferroelectrics with a nonhomogeneous distribution of the nonlinearity. For the complete characterization of the Fourier spectrum, three independent parameters must be taken into account: Two of them are the coordinates in the far-field plane, the third must relate to the mismatch $\Delta \mathbf{k}$. The latter depends on easily controlled experimental quantities such as the FF angle of incidence, wavelength and sample temperature [8], [19]. Imaging of the generated SH far field at various FF wavelengths can therefore provide information on the dominant components of the nonlinear coefficient in reciprocal space, while standard inspection techniques such as piezo-atomic force microscopy or charge-selective etching can only give information on ferroelectric domains near the sample surface. The extension of this concept to the general cases of sum and difference frequency generation by undepleted pumps is straightforward.

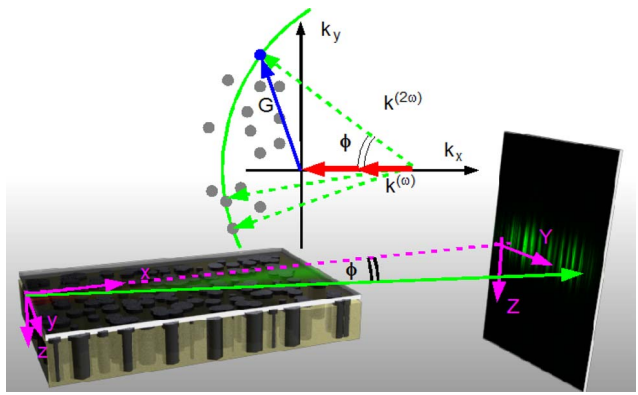

Fig. 1. Geometry. z-polarized FF pump beam propagates along *x* in a crystal with a 2-D random distribution of domains with alternate signs. The z-polarized SH is generated with a distribution related to the spatial spectrum of the nonlinear domains. $\phi$ is the angle of a generic SH plane-wave component in *xy*. The inset shows the Fourier spectrum of the nonlinear distribution, with phase mismatch versus $\phi$.

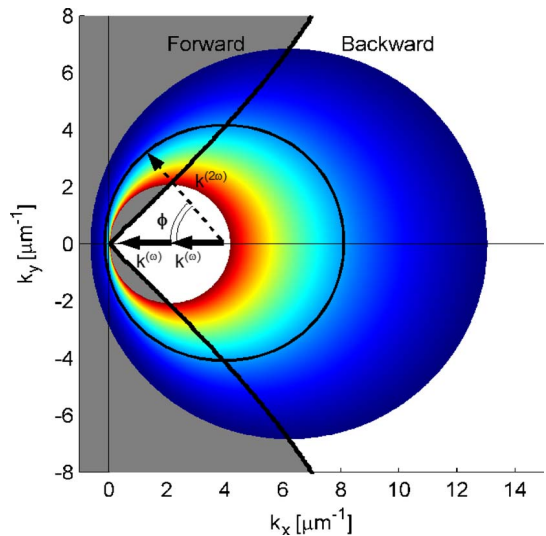

Fig. 2. Phase mismatch in the Fourier space of $LiTaO_3$ for a pump propagating along *x* versus the angle $\phi$. Gray and white areas refer to regions for forward and backward SH propagation, respectively. The wavelength is tuned from (blue circle) 700 nm to (red circle) 2 $\mu$m. Inset: zoom-in around the origin.

To gain more physical insight, we focus our attention on a uniaxial crystal with randomly distributed nonlinear domains in the principal plane *xy* where the extraordinarily polarized FF pump propagates: TWM takes place with an extraordinarily polarized SH through the nonzero quadratic tensor element $d_{33}$. [21]. The relation in Eq. (4) for this interaction can be simplified as

$$E^{(2\omega)}(Y) \propto \mathcal{F}\left\{\left[E_o^{(\omega)}(x,y)\right]^2 d_{33}(x,y)\right\}(\Delta k_x, \Delta k_y)$$
$$\Delta k_x = 2k^{(\omega)} - k^{(2\omega)}\cos(\phi)$$
$$\Delta k_y = -k^{(2\omega)}\sin(\phi) \tag{5}$$

where *Y* is the transverse coordinate on the far-field screen (see Fig. 1).

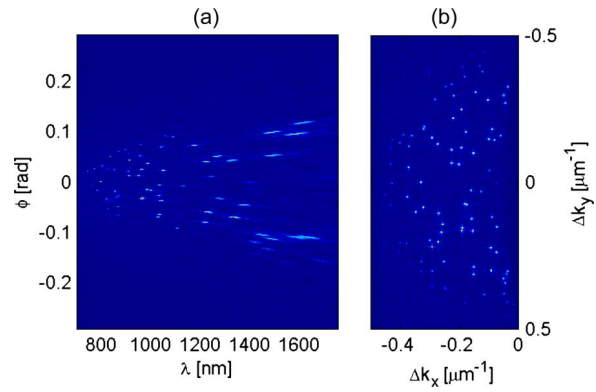

Fig. 3. Random domain mapping; numerical example for a ferroelectric crystal such as LiTaO$_3$. (a) Distribution of SH intensity versus FF wavelength and internal angle of propagation $\phi$. (b) SH intensity in Fourier space.

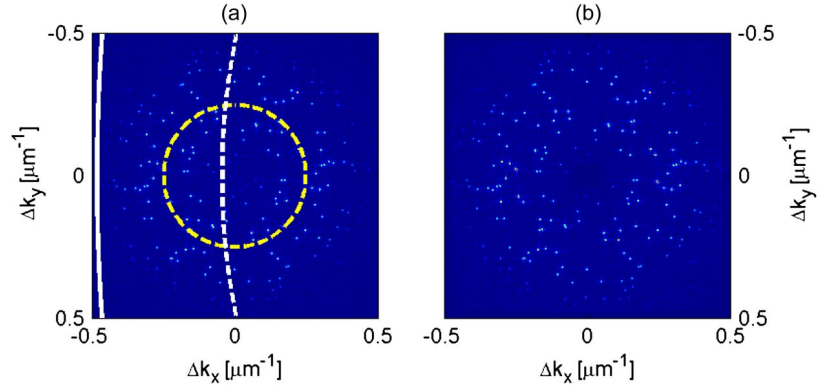

Fig. 4. Random domain mapping; numerical example. (a) Fourier transform of the nonlinear random pattern. Dashed yellow circle: Fourier components corresponding to 6-$\mu$m domain sizes. White lines: Constant wavelength loci at (solid line) 700 nm and (dashed line) 1800 nm, respectively. (b) Recovered spectrum via SHG.

The range of spatial frequencies that can be measured is defined by the phase mismatch and can be evaluated with the aid of the Sellmeier equations for the specific crystal. For example, frequencies between $2 \times 10^3$ cm$^{-1}$ and $5 \times 10^5$ cm$^{-1}$, corresponding to spatial periodicity from 110 nm to 30 $\mu$m, respectively, can be addressed by tuning the FF wavelength from 700 nm to 2000 nm in LiTaO$_3$ (see Fig. 2) and collecting the forward-propagating SH. The asymmetry of the accessible range with respect to the origin can be partially overcome by considering that the ferroelectric domain distribution is a real function; hence, its Fourier transform possesses complex conjugate symmetry with respect to the origin. If the FF angle of propagation can be continuously tuned in a solid angle, *all* frequencies in the previous domain are in principle accessible.

As an example, we consider the reciprocal spectrum of a LiTaO$_3$ ferroelectric crystal with randomly distributed domain signs in the range 2.5 $\mu$m–20 $\mu$m. Taking a Gaussian distribution centered around 6 $\mu$m and a square sample of size $0.5 \times 0.5$ mm$^2$, we solved for type-I SHG using a standard (nonlinear) beam propagator and a plane-wave FF input; then, we calculated the far-field SH after exiting the crystal and subsequently propagating in air. The final result is a map of the far-field SH in the $\lambda - \phi$ space, where $\phi$ is the internal angle of propagation obtained by Snell's law, as shown on the left side of Fig. 3. These maps represent a portion of the Fourier spectrum of the nonlinear coefficient in $\lambda - \phi$. Using Eq. (5), it is possible to convert the map in the Fourier space.

The whole spectrum is recovered by rotating the sample by 90° and assuming complex conjugate symmetry. The comparison between the original pattern of the Fourier transform modulus and the obtained SH is visible in Fig. 4. A correlation as high as 0.97 is obtained between the two diagrams.

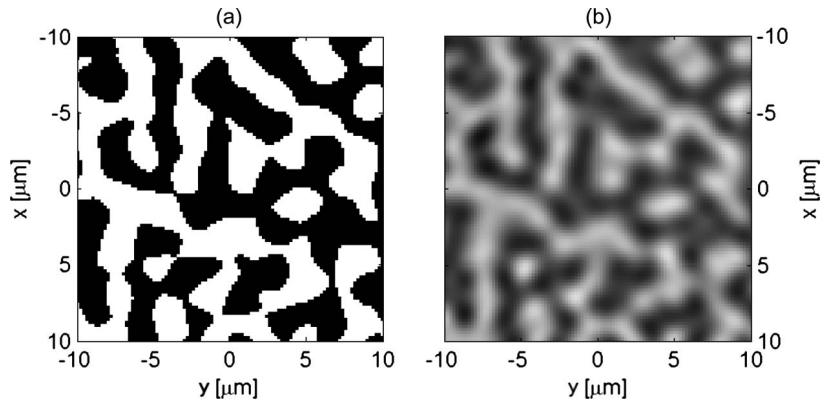

Fig. 5. Random domain mapping, numerical example. (a) Random pattern of inverted nonlinear domains. (b) Recovered pattern using far-field SHG and Fourier transform.

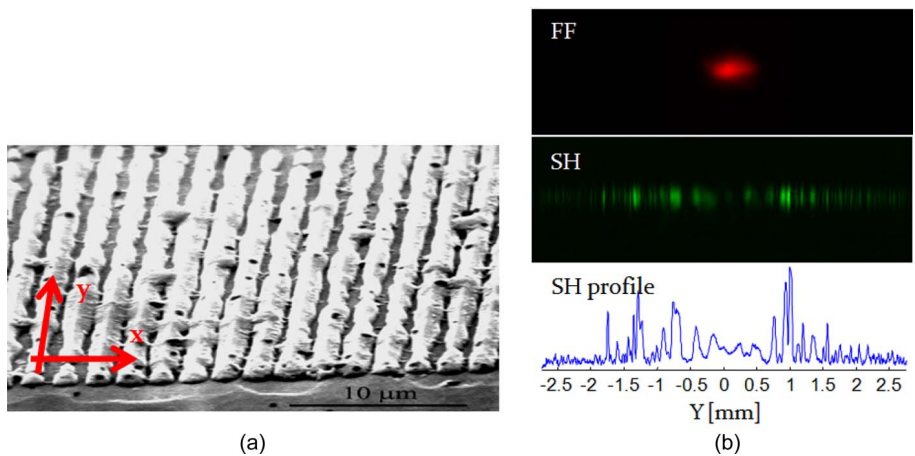

Fig. 6. Random mark-to-space ratio and SHG far-field spectroscopy. (a) Micrograph of the SPP-LT sample after chemical etching: a nonuniform mark-to-space ratio is visible in the poled area with 2-$\mu$m period. (b) Far-field profiles for the FF beam ($\lambda = 930$ nm) and for the SH. Due to the random QPM generation, the SH spreads out in the output plane.

If the SH phase were known (e.g., by interferometric measurements), then the map could be transformed back in direct space, as shown in Fig. 5. The domain borders in the recovered pattern are smoothed out due to cutoff at high frequencies.

## 3. Experiments in Lithium Tantalate

In the experiments we employed a surface periodically poled LiTaO$_3$ crystal. Surface periodic poling (SPP) is a planar technique for preparing quasi-phase-matched ferroelectrics with short periodicities [22].

The poling field is applied along the $z$-axis and inverts all the ferroelectric domains with the exclusion of those under the masked region, giving rise to uninverted domains at the surface of the sample. We used a mask with a 1-D grating of 2-$\mu$m period. As visible in Fig. 6(a), owing to the nucleation dynamics, the mark-to-space ratio after poling was nonuniform in the $(x, y)$ plane, with a random component superimposed to the QPM grating. In order to perform the nonlinear characterization of the sample, we focused a Ti:Sapphire laser beam (FF wavelength tunable from 900 to 980 nm and a $\approx$40-GHz linewidth) to a 25-$\mu$m waist, propagating it along the $x$-axis for 1 cm, using a linear $z$-polarization to achieve type-I $e-ee$ SHG using $d_{33}(x, y)$. We collected SH images at a distance d = 2 cm from the output facet using a CCD camera; the maximum value of half-waist of the FF beam inside the sample $HW \approx 15$ $\mu$m allowed us to adopt the

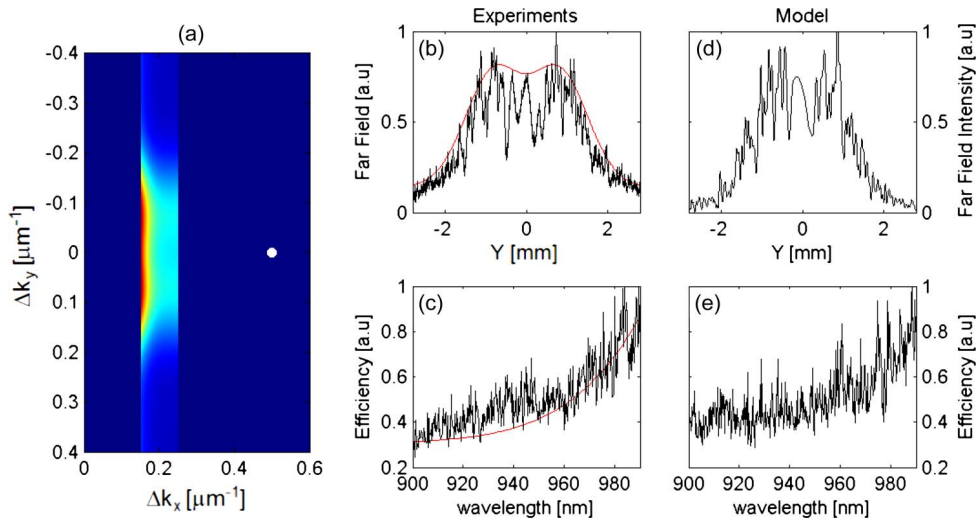

Fig. 7. Measurement of the spectral energy density. (a) Pseudo-color map for the extracted profile of the spectral density of the random sample. The dark blue region was unexplored in our experimental campaign. The 2-$\mu$m period QPM is indicated by a white dot (overscaled). Center column: Experimental data. (b) Far-field SH profile averaged over 50 measurements with FF at 930 nm (black). (c) (Black) Conversion efficiency measurements; the red lines are fits corresponding to the spectral density on the left. Right: Numerical results; the interpolated profile on the left was used to implement a Gaussian realization of the randomness in $d_{33}(x, y)$. (d) Simulation of SH far field over 20 realization. (e) Calculated conversion efficiency.

Fraunhofer approximation, i.e., $max(|\mathbf{r}' \times \hat{\mathbf{r}}|^2)/(2\lambda^{(2\omega)}) = HW^2/(2\lambda^{(2\omega)}) \approx 240$ $\mu$m, which is much smaller than the propagation distance. Fig. 6(b) shows the images of the far-field patterns for the FF and the SH. The SH pattern is essentially 1-D, confirming the homogeneity along the $z$-axis in the illuminated zone; its profile across $Y$ is plotted on the bottom of Fig. 6(b).

The Fourier transform approach can be applied to extract information on the statistic distribution of the random nucleation of domains. Let us consider the domain distribution as an aleatory process, i.e., $d_{33}$ as an aleatory variable in the $(x, y)$ plane. We can use the energy spectral density, defined as the mean of the square modulus of the Fourier transform $D(k_x, k_y)$ (i.e., $\langle |D(k_x, k_y)|^2 \rangle$) to characterize the random process.

We evaluated this parameter by averaging the far-field SH image over several positions of the FF beam in the sample, assembling an interpolating function of the energy spectral density in the $(k_x, k_y)$ space for the explored interval, as shown in Fig. 7(a). The center top of Fig. 7 plots the average of the SH far field for an FF wavelength of 930 nm. The profile is interpolated by two symmetrical Gaussians. Data collected at 900, 930 and 960 nm yield to comparable transverse profiles. The average conversion efficiency versus fundamental wavelength is the integral along the $\lambda = const$ locus of the energy spectral density. Fig. 7(c) shows the measured conversion efficiency, which increases with the FF wavelength.

The information on transverse profile and efficiency allowed for the interpolation the energy spectral density [see Fig. 7(a)] in the explored wavelength range at the FF. The profile in Fig. 7(b) indicates that noise was symmetrically distributed along $y$. Our scanning range was too narrow to appreciate variations in the parameters of the two interpolating Gaussian. Nevertheless, the reconstructed fitting function in Fig. 7(a) suggests that the distribution does not possess a radial symmetry, as it would be the case of a noise isotropically distributed in the plane. This indicates an influence of the mask for periodic poling imposed along $x$ on the (spontaneous) nucleation of random domains. In our range of observations we measured domain features comprised between 3 and 6 $\mu$m, with the largest occurrence of domains of 5.3 $\mu$m.

It is possible to mimic the random trend in the used range of wavelengths by generating the $d_{33}$ distribution as a random Gaussian process, filtered by the interpolating function in the Fourier space. We superimposed this distribution to a sinusoidal one along $x$ for the 2-$\mu$m period QPM, then we extracted the sign of $d_{33}$. In the range 900–1000 nm the statistic behavior was well reproduced, as shown in Fig. 7(a). The SH at the QPM resonance ($\approx$730 nm) was efficiently generated with a bell-shape output (as experimentally verified [22]).

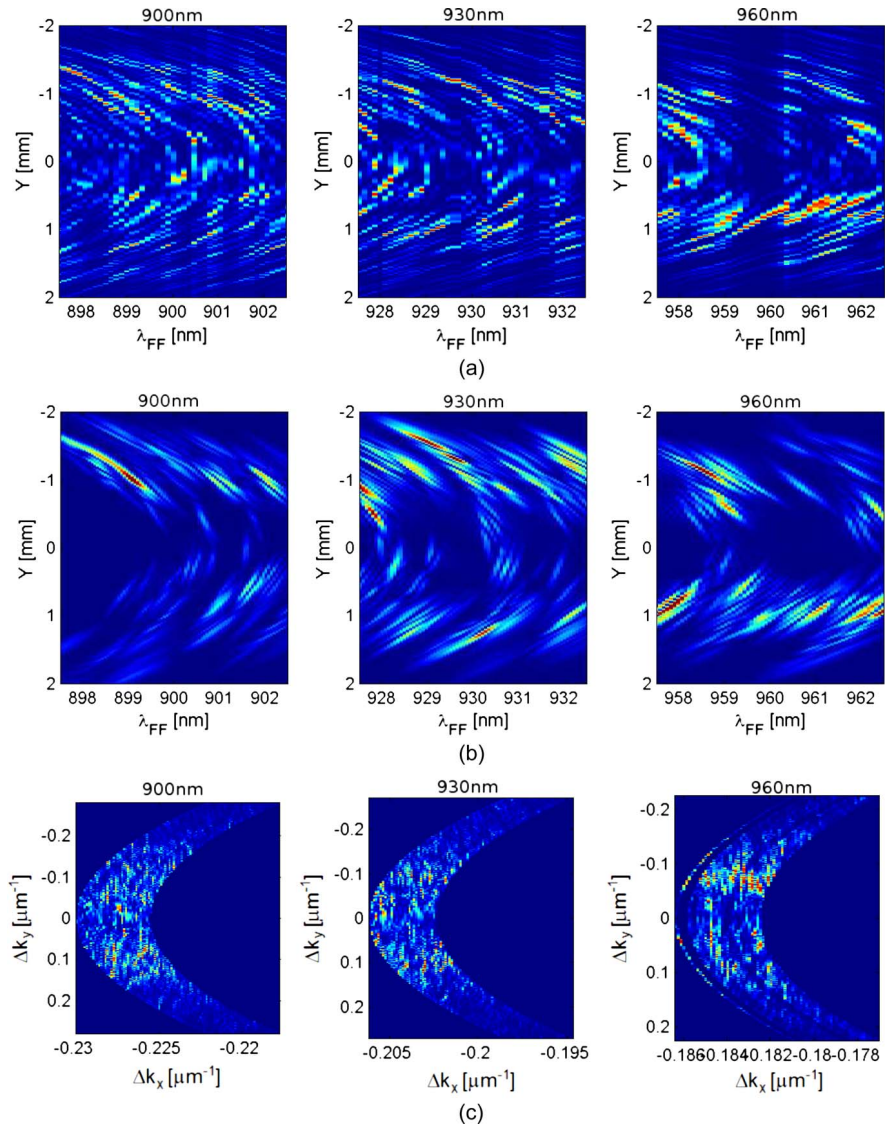

Fig. 8. Fourier spectrum measurements. (a) SH profiles versus fundamental wavelength around 900, 930, and 960 nm (left to right), respectively. The loci $\Delta k_x = constant$ are clearly visible. (b) Numerically evaluated SH profiles versus fundamental wavelength around 900, 930, and 960 nm (left to right). (c) Fourier transform obtained from Fig. 8(a) using Eq. (5). The SH profiles are shown in the Fourier coordinates $\Delta k_x$, $\Delta k_y$.

By focusing our attention on a specific region of the sample, it is possible to analyze specific details of the Fourier spectrum. The narrow laser linewidth enabled us to fine-tune the phase mismatch $\Delta k$ and collect several images which were rearranged in the maps in Fig. 8(a), with a fundamental wavelength span of 6 nm around 900, 930, and 960 nm, respectively.

The loci $\Delta k_x = const$ are clearly visible in the map, as paths varying along $Y$ with concavity toward shorter wavelength. $E_o(x, y)$ has a strong asymmetry in $x, y$, since it is limited in $y$ to tens of micrometers while in $x$ it covers the whole sample length of 1 cm. In Fourier space, the convolution between the Fourier transform of the fundamental profile $[E_o(x,y)]^2$ and the distribution $d_{33}$ results in the blurring of the $d_{33}$ transform in $\Delta k_y$. Even with this limitation, the regions of higher intensity point out the dominant spatial frequency contributions to the nonlinear coefficient. To numerically mimic this behavior we adjusted the Fourier transform of $d_{33}(x,y)$ to exhibit the same peak of the reconstructed space in Fig. 8(c). The resulting Fig. 8(b) shows a good agreement with the data of Fig. 8(a) in terms of SH energy distribution.

## 4. Conclusion

In conclusion, we introduced an accessible, noninvasive approach to characterize ferroelectric domain distribution in noncentrosymmetric crystals; the approach, based on SHG far-field spectroscopy, demonstrates that it is possible to extract information on the Fourier spectrum by scanning a convenient set of FF wavelengths. We experimentally tested this procedure using a surface periodically poled LiTaO$_3$ sample with nonuniform mark-to-space ratio and managed to extract global statistical parameters such as the energy spectral density of the random domain distribution. From the obtained information we could numerically reproduce the experimental results. These findings demonstrate the potential in terms of diagnostics and morphological analysis afforded in ferroelectrics by a TWM when matched to Fourier optics, paving the way to applications of random domain poling toward the engineering of versatile and broadband wavelength converters.

## Appendix
## Derivation in the Presence of Anisotropy

The scalar approximation can be reductive when dealing with propagation in anisotropic media such as nonlinear crystals. We derive hereby a relation accounting for both ordinary- and extraordinary-wave propagation in uniaxial media with optic axis along $z$ and refractive indices $n_o$ and $n_e$, respectively. We start with the Maxwell's equations for the SH, considering an external source $\mathbf{P}^{NL}(\mathbf{r}) = \underline{\underline{\chi}}^{(2)}(\mathbf{r}) : \mathbf{E}^{(\omega)}\mathbf{E}^{(\omega)}$, where the nonlinear susceptibility $\underline{\underline{\chi}}^{(2)}$ is a third-order tensor. We omit the superscript $(2\omega)$ in the following:

$$\nabla \times \mathbf{E} = i\omega\mu_0 \mathbf{H}$$
$$\nabla \times \mathbf{H} = -i\omega\epsilon_0(\underline{\underline{\epsilon}}\mathbf{E} + \mathbf{P}^{NL}) \quad (6)$$

or, equivalently

$$\nabla^2 \mathbf{E} - \nabla\nabla \cdot \mathbf{E} + k_0^2 \underline{\underline{\epsilon}}\mathbf{E} = -k_0^2 \mathbf{P}^{NL}$$
$$\nabla \cdot (\underline{\underline{\epsilon}}\mathbf{E}) = n_o^2 \nabla_\perp \cdot \mathbf{E}_\perp + n_e^2 \partial_z E_z = -\nabla \cdot \mathbf{P}^{NL} \quad (7)$$

where the dielectric constant $\underline{\underline{\epsilon}}$ is a second-order tensor, diagonal in the reference system with principal diagonal $\{n_o^2, n_o^2, n_e^2\}$. The two equations above can be cast in the form

$$\mathcal{L}(\mathbf{E}) = -k_0^2 \mathbf{P}^{NL} - \frac{1}{n_o^2 n_e^2} \underline{\underline{\epsilon}}\nabla\nabla \cdot \mathbf{P}^{NL} \quad (8)$$

with

$$\mathcal{L} = \begin{pmatrix} \frac{n_o^2}{n_e^2}\partial_{xx} + \partial_{yy} + \partial_{zz} + n_o^2 k_0^2 & -\frac{(n_e^2-n_o^2)}{n_e^2}\partial_{xy} & 0 \\ -\frac{(n_e^2-n_o^2)}{n_e^2}\partial_{xy} & \partial_{xx} + \frac{n_o^2}{n_e^2}\partial_{yy} + \partial_{zz} + n_o^2 k_0^2 & 0 \\ 0 & 0 & \partial_{xx} + \partial_{yy} + \frac{n_e^2}{n_o^2}\partial_{zz} + n_e^2 k_0^2 \end{pmatrix}. \quad (9)$$

Defining the vector potential **A**

$$\mathbf{E} = -k_0^2 \mathbf{A} - \frac{1}{n_o^2 n_e^2} \boldsymbol{\epsilon} \nabla \nabla \cdot \mathbf{A} \tag{10}$$

the system reduces to

$$\mathcal{L}\mathbf{A} = \mathbf{P}^{NL}. \tag{11}$$

We can define the linear scalar operators $\mathcal{L}_o(\mathcal{L}_e)$ ruling ordinary-wave (extraordinary-wave) propagation and possessing the Green functions $\mathbf{G}_o$ ($\mathbf{G}_e$) as [23]

$$\mathcal{L}_o = \nabla^2 + k_0^2 n_o^2, \qquad G_o(\mathbf{r})^{(\pm)} = \frac{e^{\pm i k_0 n_o |\mathbf{r}|}}{4\pi |\mathbf{r}|} \tag{12}$$

$$\mathcal{L}_e = \partial_{xx} + \partial_{yy} + \frac{n_e^2}{n_o^2} \partial_{zz} + k_0^2 n_e^2, \qquad G_e(\mathbf{r})^{(\pm)} = \frac{n_o e^{\pm i k_0 n(\theta) |\mathbf{r}|}}{4\pi n(\theta) |\mathbf{r}|} \tag{13}$$

where $n(\theta)^2 = n_o^2 \cos^2(\theta) + n_e^2 \sin^2(\theta)$ is the extraordinary-wave refractive index, depending on the polar angle $\theta$ on the polar axis $z$. [23] System (11) can be expressed as a function of the two previous operators as

$$\mathcal{L}_o(\nabla \times \mathbf{A}_\perp \cdot \hat{\mathbf{z}}) = \nabla \times \mathbf{P}_\perp^{NL} \cdot \hat{\mathbf{z}}$$
$$\mathcal{L}_e(\nabla_\perp \cdot \mathbf{A}_\perp) = \nabla_\perp \cdot \mathbf{P}_\perp^{NL}$$
$$\mathcal{L}_e(A_z) = P_z^{NL}. \tag{14}$$

We split the contribution of the nonlinear polarization (i.e., the source generating the SH) in two components $P_z^{NL}$ and $\mathbf{P}_\perp^{NL}$ parallel and orthogonal to the optic axis $z$, respectively. The former always generates extraordinary waves, while the second can generate both ordinary and extraordinary waves. Note that the Helmoltz theorem allows decomposing the field $\mathbf{P}_\perp^{NL}$ (and then $\mathbf{A}_\perp$) in the sum of a solenoidal (divergence-free $\mathbf{P}_{\perp df}^{NL}$) and an irrotational(curl-free $\mathbf{P}_{\perp cf}^{NL}$) vector fields in the plane $xy$. The solutions can be obtained using the Green functions defined in (13):

$$\mathcal{L}_o(\mathbf{A}_{\perp df}) = \mathbf{P}_{df}^{NL}, \quad \mathbf{A}_{\perp df} = \int G_o(\mathbf{r} - \mathbf{r}')^{(+)} \mathbf{P}_{\perp df}^{NL}(\mathbf{r}') d\mathbf{r}' \tag{15}$$

$$\mathcal{L}_e(\mathbf{A}_{\perp cf}) = \mathbf{P}_{\perp cf}^{NL}, \quad \mathbf{A}_{\perp cf} = \int G_e(\mathbf{r} - \mathbf{r}')^{(+)} \mathbf{P}_{\perp cf}^{NL}(\mathbf{r}') d\mathbf{r}' \tag{16}$$

$$\mathcal{L}_e(A_z) = P_z^{NL}, \quad A_z = \int G_e(\mathbf{r} - \mathbf{r}')^{(+)} P_z^{NL}(\mathbf{r}') d\mathbf{r}'. \tag{17}$$

Employing transformation (10) we retrieve the electric fields

$$\mathbf{E}_o = -k_0^2 \int G_o(\mathbf{r} - \mathbf{r}')^{(+)} \mathbf{P}_{\perp df}^{NL}(\mathbf{r}') d\mathbf{r}'$$

$$\mathbf{E}_e = -\left(k_0^2 + \frac{1}{n_o^2 n_e^2} \boldsymbol{\epsilon} \nabla \nabla \cdot \right) \int G_e(\mathbf{r} - \mathbf{r}')^{(+)} \left( \mathbf{P}_{\perp cf}^{NL}(\mathbf{r}') + P_z^{NL} \hat{\mathbf{z}} \right) d\mathbf{r}'. \tag{18}$$

The extraordinary and ordinary components in the far field (neglecting refractive index discontinuities) can be expressed in terms of the Fourier transform of the nonlinear polarization $\mathbf{P}^{NL}(\mathbf{r}) = \mathbf{P}_0^{NL}(\mathbf{r})e^{2i\mathbf{k}^{(\omega)}} = \underline{\underline{\chi}}^{(2)}(\mathbf{r}) : \mathbf{E}^{(\omega)}\mathbf{E}^{(\omega)}$.

$$\mathbf{E}_o^{(2\omega)} = -k_0^2 \frac{e^{ik_0^{(2\omega)} n_o^{(2\omega)} r}}{2r} \mathcal{F}\left\{\mathbf{P}_0^{NL}(\mathbf{r})_{\perp df}\right\}\left(2\mathbf{k}^{(\omega)} - \mathbf{k}_0^{(2\omega)} n_o^{(2\omega)}\right)$$

$$\mathbf{E}_e^{(2\omega)} = -\left(k_0^2 + \frac{1}{n_o^2 n_e^2}\boldsymbol{\varepsilon}\nabla\nabla\cdot\right)\frac{n_o^{(2\omega)} e^{ik_0^{(2\omega)} n(\theta)^{(2\omega)} r}}{2n(\theta)^{(2\omega)} r}$$
$$\times \mathcal{F}\left\{\mathbf{P}_0^{NL}(\mathbf{r})_{\perp cf} + P_0^{NL}(\mathbf{r})_z \hat{z}\right\}\left(2\mathbf{k}^{(\omega)} - \mathbf{k}_e^{(2\omega)}\right) \quad (19)$$

with

$$\mathbf{k}_e = k_0 \frac{(n_e^2 \sin\theta(\cos\phi\hat{x} + \sin\phi\hat{y}) + n_o^2 \cos\theta\hat{z})}{n(\theta)}.$$

## Acknowledgment

We are grateful to Prof. S. Riva Sanseverino for encouragement and project coordination.